\documentclass[twocolumn,showpacs,superscriptaddress,prl]{revtex4}
\usepackage{graphicx}
\usepackage{epsfig}
\usepackage{color}
\usepackage{amsmath,amssymb,subfigure,psfrag}
\def\U#1{{\rm #1}} 
\def\u#1{_{\rm #1}}

\newcommand{\bra}[1]{\langle #1 |}
\newcommand{\ket}[1]{| #1 \rangle}

\newcommand{\dagg}[1]{#1 ^\dagger}

\def\Dv{\U{D}\u{V}}

\def\Da{\U{D}\u{T1}}
\def\Db{\U{D}\u{T2}}

\begin{document}
\title{
Non-classical two-photon interference 
between independent telecom light pulses
converted by difference-frequency generation
}

\author{Rikizo~Ikuta}
\affiliation{Graduate School of Engineering Science, Osaka University,
Toyonaka, Osaka 560-8531, Japan}
\author{Toshiki~Kobayashi}
\affiliation{Graduate School of Engineering Science, Osaka University,
Toyonaka, Osaka 560-8531, Japan}
\author{Hiroshi~Kato}
\affiliation{Graduate School of Engineering Science, Osaka University,
Toyonaka, Osaka 560-8531, Japan}
\author{Shigehito~Miki}
\affiliation{
Advanced ICT Research Institute, 
National Institute of Information and Communications Technology (NICT), 
Kobe 651-2492, Japan}
\author{Taro~Yamashita}
\affiliation{
Advanced ICT Research Institute, 
National Institute of Information and Communications Technology (NICT), 
Kobe 651-2492, Japan}
\author{Hirotaka~Terai}
\affiliation{
Advanced ICT Research Institute, 
National Institute of Information and Communications Technology (NICT), 
Kobe 651-2492, Japan}
\author{Mikio~Fujiwara}
\affiliation{
Advanced ICT Research Institute, 
National Institute of Information and Communications Technology (NICT), 
Koganei, Tokyo 184-8795, Japan}
\author{Takashi~Yamamoto}
\affiliation{Graduate School of Engineering Science, Osaka University,
Toyonaka, Osaka 560-8531, Japan}
\author{Masato~Koashi}
\affiliation{Photon Science Center, The University of Tokyo, 
Bunkyo-ku, 113-8656, Japan}
\author{Masahide~Sasaki}
\affiliation{
Advanced ICT Research Institute, 
National Institute of Information and Communications Technology (NICT), 
Koganei, Tokyo 184-8795, Japan}
\author{Zhen~Wang}
\affiliation{
Advanced ICT Research Institute, 
National Institute of Information and Communications Technology (NICT), 
Kobe 651-2492, Japan}
\author{Nobuyuki~Imoto}
\affiliation{Graduate School of Engineering Science, Osaka University,
Toyonaka, Osaka 560-8531, Japan}

\begin{abstract}
We experimentally demonstrated 
the Hong-Ou-Mandel~(HOM) interference between two photons 
after visible-to-telecommunication wavelength conversion. 
In the experiment, 
we prepared a heralded single photon 
by using spontaneous parametric down-conversion 
and the other photon from a weak laser source at 780 nm. 
We converted the wavelength of both photons 
to the telecommunication wavelength of 1522 nm 
by using difference-frequency generation 
in a periodically-poled lithium niobate, 
and then observed the HOM interference between the photons. 
The observed visibility is $0.76\pm 0.12$ 
which clearly 
shows the non-classical interference of the two photons. 
The high-visibility interference is an important step 
for fiber-based quantum communications of photons 
generated from visible photon emitters. 
\end{abstract}

\maketitle
Quantum interface for wavelength conversion 
of photons~\cite{conv} 
is an important building block for linking quantum information 
among different kinds of physical systems, 
and it has been actively studied~\cite{tanzilli, takesue, 
matthew, mcguinness, dudin, ikuta, ramelow, ikuta2, ates, zaske}. 
Especially, for optical-fiber-based quantum communication 
over a long distance with quantum repeaters~\cite{repeaters}, 
wavelength conversion of photons 
entangled with quantum memories to telecommunication bands 
without destroying their quantum information is required. 
Many promising quantum memories have resonant wavelengths 
at the visible range~\cite{matsukevich, rosenfeld, 
chou, olmschenk, togan, ritter, hofmann}, 
which necessitates a quantum interface 
for visible-to-telecommunication wavelength conversion. 
\begin{figure}[t]
 \begin{center}
  \scalebox{0.58}{\includegraphics{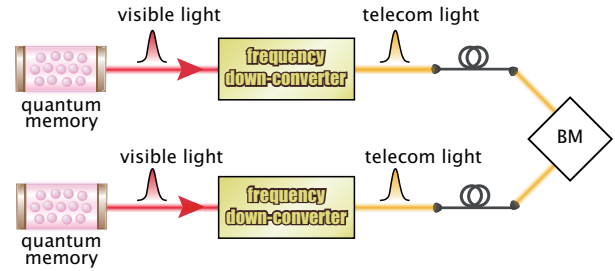}}
  \caption{
  An elementary link of quantum repeaters 
  with the quantum interface. 
  Each quantum memory has two excitation modes 
  coupled to two modes of a photon such as polarization. 
  Visible photons entangled with quantum memories 
  are frequency down-converted to the telecommunication bands, 
  and pass through optical fibers. 
  The Bell measurement is performed on the two photons 
  for establishing entanglement between the quantum memories. 
  \label{fig:concept}}
 \end{center}
\end{figure}
Furthermore, in order to establish an entanglement 
between the remote quantum memories 
through two-photon interference~\cite{hofmann, beugnon, 
maunz, moehring, bernien, sipahigil}, 
two down-converted telecom photons should be detected 
by the Bell measurement~(BM) at the intermediate relay node, 
as illustrated in Fig.~\ref{fig:concept}. 
In the BM, the two telecom photons must be indistinguishable, 
and cause non-classical interference, 
i.e. the Hong-Ou-Mandel~(HOM) interference~\cite{HOM}. 
Entanglement-preserving visible-to-telecommunication 
wavelength conversions have been demonstrated 
in~\cite{dudin, ikuta, ikuta2}, and highly entangled states 
have been observed~\cite{dudin, ikuta2}. 
However, the HOM interference at the relay node 
has never been demonstrated. 
Related experiments have been 
restricted to the HOM interference 
between up-converted visible photons~\cite{takesue, ates}. 

In this Letter, 
we present the first demonstration of the HOM interference 
between two light pulses at the telecommunication band 
after frequency down-conversion, 
which matches up with fiber-based quantum communication 
with quantum repeaters. 
We initially prepared two pulses at 780 nm, 
one being a heralded single photon 
generated from spontaneous parametric down-conversion~(SPDC) 
and the other being a coherent light pulse directly from the laser. 
Their wavelengths were converted to 1522 nm 
by difference-frequency generation~(DFG) 
using a periodically-poled LiNbO$_3$~(PPLN) waveguide. 
We then observed the HOM interference 
between the converted light pulses. 
The interference visibility was $0.76\pm 0.12$, 
which clearly exceeds the maximum value of 0.5 
in the classical wave theory~\cite{Ou}. 
A comparison to our theoretical analysis shows 
that the degradation of the visibility is 
mainly caused by the multi-photon events 
from the coherent light pulse, 
which suggests that 
the demonstrated frequency down-converter will enable us 
to achieve a visibility as high as 0.91. 

\begin{figure}[t]
 \begin{center}
  \scalebox{0.58}{\includegraphics{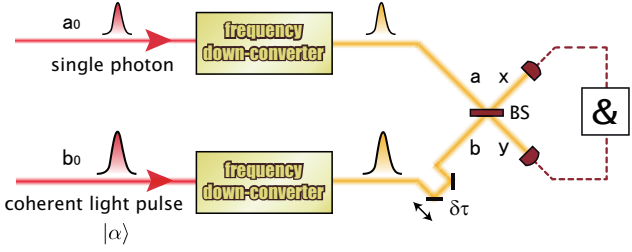}}
  \caption{
  Concept of our experimental setup. 
  A single photon is prepared 
  by measuring one of the photon pair from the SPDC. 
  The coherent light pulse is prepared by the laser source. 
  \label{fig:concept2}}
 \end{center}
\end{figure}
A conceptual setup of 
the HOM interference after the wavelength conversion in our experiment 
is shown in Fig.~\ref{fig:concept2}. 
The two input visible light pulses are 
a coherent light pulse and a heralded single photon. 
The two pulses have no correlations and 
they are regarded as prepared 
from independent sources~\cite{Rarity1, Rarity2, Ruibo}. 
The two visible light pulses are frequency down-converted 
to the telecommunication wavelength, 
and then they are combined by a half beamsplitter~(BS). 
When the photons from the two pulses are indistinguishable, 
they always leave the same output port of the BS. 
On the other hand, 
the photons may appear in both output ports 
if these are not completely indistinguishable. 
By measuring the coincidence photon detection 
between the two output modes of the BS while changing 
the time delay $\delta \tau$ between the two light pulses, 
we observe the HOM dip. 
In the classical wave theory, 
the maximum of the visibility of the HOM interference is $0.5$. 

\begin{figure}
 \begin{center}
  \scalebox{0.47}{\includegraphics{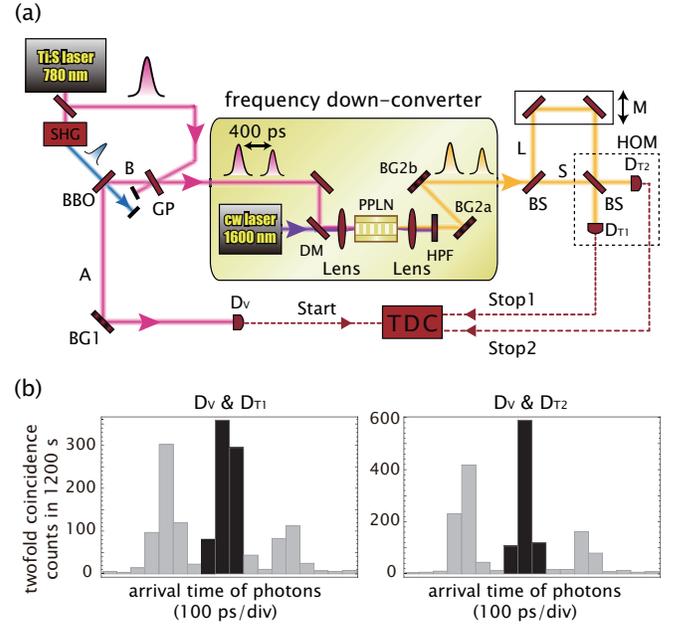}}
  \caption{
  (a)
  The experimental setup for the HOM interference 
  after the wavelength conversion. 
  The single PPLN waveguide is used 
  for the wavelength conversion of the heralded single photon 
  and the coherent light pulse. 
  (b)
  Two-fold coincidence counts between $\Dv$ and $\Da$, 
  and between $\Dv$ and $\Db$. 
  In each histogram, 
  three peaks of the arrival time of the stop signal are observed. 
  For the threefold coincidence events among $\Dv$, $\Da$ and $\Db$, 
  we post-select the 300-ps time windows of the two central peaks. 
  \label{fig:setup}}
 \end{center}
\end{figure}
The experimental setup for the HOM interference 
is shown in Fig.~\ref{fig:setup}~(a). 
A pico-second light pulse 
from a mode-locked Ti:sapphire~(Ti:S) laser~(wavelength: 780 nm; 
pulse width: $\Delta t\equiv $1.2 ps; repetition rate: 82 MHz) 
is divided into two beams. 
One beam is frequency doubled~(wavelength: 390 nm; power: 200 mW) 
by second harmonic generation~(SHG), 
and then pumps a Type-I phase-matched 1.5mm-thick 
$\beta$-barium borate~(BBO) crystal 
to generate a vertically(V)-polarized photon pair in modes A and B 
through SPDC. 
The photon in mode A is measured 
by a superconducting single photon detector~(SSPD)~\cite{NICT1, NICT2} 
denoted by $\Dv$ for preparing a heralded single photon in mode B. 
The spectral filtering of the photon in mode A is 
performed by a Bragg grating~(BG1) with a bandwidth of 
$\Delta\u{BG1}\equiv $0.2 nm. 
The other beam is weakened to an average photon number of 
$\sim 0.5$ by using a glass plate~(GP), 
and is sent along the same light path as the photon B. 
Time difference 
between the photon B and 
the subsequent coherent light pulse is set to be $\sim 400$ ps 
which is 
longer than the temporal resolution $150$ ps of the detectors~\cite{ikuta2}. 
The two light pulses in the visible range 
are then sent to the frequency down-converter. 

In the frequency down-converter, 
a V-polarized cw pump laser at 1600 nm 
with an effective power of 500 mW 
is combined with the signal beams at 780 nm 
by a dichroic mirror~(DM). 
They are focused on the Type-0 quasi-phase 
matched PPLN waveguide~\cite{nishikawa}. 
The length of the PPLN crystal is 20 mm and 
the acceptable bandwidth is calculated to be 
$\Delta\u{WG}\equiv $0.3 nm. 
After passing through the PPLN waveguide, 
the strong pump light is reduced 
by a high-pass filter~(HPF), 
and the light converted to the wavelength of 1522 nm 
is extracted by BG2a and BG2b with bandwidths of 
$\Delta\u{BG2}\equiv $ 1 nm. 

The light pulses from the frequency down-converter are 
split into a short path~(S) and a long path~(L) by a BS. 
Time difference between S and L 
is changed by mirrors~(M) on a motorized stage. 
The pulses passing through S and L are mixed by a second BS 
for the HOM interference. 
The two output beams from the BS 
are coupled to single-mode fibers 
followed by two SSPDs $\Da$ and $\Db$. 

An electric signal from $\Dv$ is 
connected to a time-to-digital converter~(TDC) 
as a start signal of a clock, 
and electric signals from $\Da$ and $\Db$ 
are connected to the TDC as stop signals. 
Fig.~\ref{fig:setup}~(b) shows the histograms 
of the delayed coincidence counts of $\Dv$\&$\Da$ and $\Dv$\&$\Db$. 
The central peak among the observed three peaks 
in each histogram includes the events 
where two photons simultaneously arrived at the BS, 
i.e., the heralded single photon passed through 
path L and the coherent light pulse passed through path S. 
Therefore, to see the HOM interference, 
we collect the coincidence events within 300-ps time windows 
of the two central peaks, 
which corresponds to the threefold coincidence events 
among $\Dv$, $\Da$ and $\Db$. 
We note that 
while the duration of the converted light pulse is $\sim 5$ ps, 
the observed widths of the peaks 
shown in Fig.~\ref{fig:setup}~(b) are much larger, 
coming from the timing resolution of our measurement system. 
The 300-ps time window was chosen such that the tails 
of the left and right peaks are negligible.

\begin{figure}[t]
 \begin{center}
 \scalebox{0.6}{\includegraphics{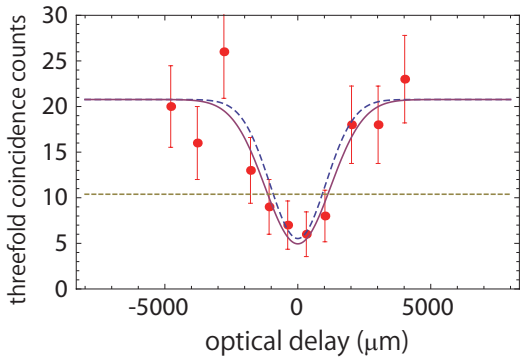}}
  \caption{
  Observed HOM interference between two light pulses. 
  The threefold coincidence counts have been observed 
  in the 300-ps time windows. 
  Solid curves are Gaussian fitted to the experimental counts. 
  Dashed curves are obtained by Eq.~(\ref{Ptau}) 
  with experimental parameters. 
  Dashed horizontal lines describe 
  the minimum values of the dips in the classical wave theory. 
  \label{fig:dip}}
 \end{center}
\end{figure}
The experimental result of the dependency 
of the threefold coincidence counts on the optical delay 
is shown in Fig.~\ref{fig:dip}, 
which clearly indicates the HOM dip. 
Each point was recorded for 23 hours. 
The observed visibility of $0.76\pm 0.12$ at the zero delay point 
was obtained by the best fit to the experimental data with a Gaussian. 
The full width at half maximum was approximately $2.9$ mm 
which corresponds to 
$\sim 10$ ps for a delay time. 
The high visibility clearly shows the non-classical interference 
of the two light pulses converted to the telecommunication band. 

To see the reasons for the degradation of the visibility, 
we construct a theoretical model 
which takes into account the mode mismatch 
between the two pulses, 
and vacuum and multi-photon components in the coherent light pulse. 
The emission rate of the photon pair from the SPDC is 
so small that the multiple-pair events are negligible 
in our experiment. 
In addition, the heralded single photon is assumed to 
be in a pure state because the narrow-band spectral filtering 
destroys the spectral correlation between the photon pairs from SPDC. 
Under the assumptions, 
the initial state composed of 
the single photon in mode $a_0$ 
and the light in the coherent state in mode $b_0$, 
which are shown in Fig.~\ref{fig:concept2}, 
is regarded as a pure state described by 
$\dagg{\hat{a}_{a_0}}\hat{D}_{b_0}(\alpha)\ket{0}$, 
where $\ket{0}$ is 
the vacuum state for both paths, 
$\dagg{\hat{a}}$ is a bosonic creation operator 
and 
$\hat{D}_{b_0}(\alpha)\equiv 
\exp(\alpha\dagg{a}_{b_0}-\alpha^*\hat{a}_{b_0})$ 
with complex number $\alpha$ is a displacement operator. 
The subscripts denote the modes 
on which the operators work. 

In the experiment for the HOM interference, 
the two light pulses are mixed at the BS and detected 
by threshold photon detectors 
which cannot distinguish the photon numbers 
and which have the same quantum efficiency 
denoted by $\eta$. 
Such a measurement is equivalent to 
photon detections with unit efficiency 
after the lights pass through a lossy channel 
with transmittance $\eta$ and 
then they are mixed by the BS~\cite{koashi}. 
As a result, 
by using effective overall transmittance $T$ which 
includes all of the photonic losses in the circuit 
such as the conversion efficiency 
of the frequency down-converter 
and the quantum efficiency of the detector, 
the input state of the light pulses in modes $a$ and $b$ 
to the BS is regarded as 
$\ket{\phi\u{T}}\equiv \dagg{\hat{a}}_a\hat{D}_{b}(\sqrt{T}\alpha)\ket{0}$ 
with probability $T$ 
and as 
$\ket{\phi\u{R}}\equiv \hat{D}_{b}(\sqrt{T}\alpha)\ket{0}$ 
with probability $1-T$. 

Two light pulses in modes $a$ and $b$ are mixed at the BS 
and channeled into paths $x$ and $y$. 
Suppose that mode $a$ is transformed 
into modes $x_a$ and $y_a$, and mode $b$ is transformed 
into modes $x_b$ and $y_b$. 
The overlap between modes 
$x_a$ and $x_b$~($y_a$ and $y_b$) 
represented by
$V \equiv |[\hat{a}_{x_a}, \dagg{\hat{a}}_{x_b}]|^2
=|[\hat{a}_{y_a}, \dagg{\hat{a}}_{y_b}]|^2$ 
depends on the time delay $\delta\tau$, 
and it may not be unity even for $\delta\tau = 0$ 
due to the mismatch in the spectral shapes of the pulses. 
When the normalized spectral amplitudes 
of modes $a$ and $b$ are assumed to be Gaussian 
with variances of $\delta\omega\u{p}^2$ and $\delta\omega\u{w}^2$ 
respectively, the overlap $V$ is given by 
\begin{eqnarray}
V(\delta \tau)
&=& \left|
\frac{1}{\sqrt{\pi \delta\omega\u{p}\delta\omega\u{w}}}
\int 
e^{-i \omega\delta \tau}
e^{-\frac{\omega^2}{2\delta\omega\u{p}^2}}
e^{-\frac{\omega^2}{2\delta\omega\u{w}^2}}
\U{d}\omega
\right|^2\\
&=&
\frac{2\delta \omega\u{p} \delta\omega\u{w}}
{\delta\omega\u{p}^2+\delta\omega\u{w}^2}
\exp\left[
-\frac{\delta\omega\u{p}^2\delta\omega\u{w}^2\delta\tau^2}
{\delta\omega\u{p}^2+\delta\omega\u{w}^2}
\right]. 
\label{Vw}
\end{eqnarray}
In our experiment, 
by using $\Delta t=$1.2~ps, 
$\Delta\u{BG1}=$0.2~nm, $\Delta\u{WG}=$0.3~nm and $\Delta\u{BG2}=$1~nm 
which are the full widths at the half maxima, 
and by assuming that all spectral functions are Gaussian, 
$V(0)$ is calculated to be larger than $0.99$. 

We also take into account the effect of background noises 
which include the Raman scattering of the cw pump light 
and the dark counts of the detectors. 
While the optical noise caused by the Raman scattering 
may interfere with the pico-second signal photons, 
such contribution is negligible in the 300-ps time window. 
We thus model the effect of these noises as a single parameter $d$, 
such that each of detectors signals a detection with probability $d$ 
even when no photons are incident. 
The twofold coincidence probability between the light pulses 
after the BS is described by 
\begin{eqnarray}
P(\delta\tau) = T P\u{T}(\delta\tau)+ (1-T)P\u{R}, 
\end{eqnarray}
where 
$P\u{T}(\delta\tau) \equiv
1-(1-d)(||\bra{0}_{x}
\hat{U}\ket{\phi\u{T}}||^2
+||\bra{0}_{y}\hat{U}\ket{\phi\u{T}}||^2)$ 
and 
$P\u{R}
\equiv 1 - 
(1-d)(||\bra{0}_{x}\hat{U}\ket{\phi\u{R}}||^2
+||\bra{0}_{y}\hat{U}\ket{\phi\u{R}}||^2
-(1-d)|\bra{0}_{x}\bra{0}_{y}\hat{U}\ket{\phi\u{R}}|^2)
$. 
$\ket{0}_i$ represents the vacuum state of path $i$. 
$\hat{U}$ is a unitary operator 
for the action of the BS, 
and it satisfies 
$\hat{U}\dagg{\hat{a}}_a \dagg{\hat{U}} 
=(\dagg{\hat{a}}_{x_a}-\dagg{\hat{a}}_{y_a})/\sqrt{2}$, 
$\hat{U}\hat{D}_b(\sqrt{T}\alpha)\dagg{\hat{U}} 
=\hat{D}_{x_b}(\sqrt{T/2}\alpha)
\hat{D}_{y_b}(\sqrt{T/2}\alpha)$ 
and $\hat{U}\ket{0}=\ket{0}_x\ket{0}_y$. 
By using 
$|\bra{0}_i\hat{D}_{i_b}(\sqrt{T/2}\alpha)\ket{0}_i|^2=\exp(-T|\alpha|^2/2)$ 
and 
$\bra{0}_i\dagg{\hat{D}}_{i_b}(\sqrt{T/2}\alpha)
\hat{a}_{i_a}\dagg{\hat{a}}_{i_a}\hat{D}_{i_b}(\sqrt{T/2}\alpha)\ket{0}_i
=V(\delta\tau) T|\alpha|^2/2+1$ for $i=x, y$, 
we have 
\begin{eqnarray}
P(\delta \tau) \propto 
1 - \frac{V(\delta \tau)}
{
\left( 1+\frac{2d}{T|\alpha|^2}
\right)
\left(
1+\frac{d}{T}+\frac{|\alpha|^2}{2}
\right)
},
\label{Ptau}
\end{eqnarray}
under the condition that 
$T\ll 1$ and $d\ll 1$ are satisfied. 
From Eq.~(\ref{Ptau}), 
we see that 
the denominator consists of two factors 
originating from 
the vacuum component and 
the multi-photon components in the coherent state. 
As a result, there is an optimal value of $|\alpha|^2$ 
for minimizing $P(0)$. 
Using Eq.~(\ref{Ptau}) and the observed parameters 
as $T\approx 0.0008$ and $d\approx 1.9\times 10^{-5}$, 
the optimal value becomes $|\alpha|^2\approx 0.31$. 
In our experiment, we chose $|\alpha|^2\approx 0.43$ 
which 
implies that the contribution from the multi-photons is larger. 

By using our experimental parameters, 
we plotted a curve predicted by 
the theoretical model written in Eq.~(\ref{Ptau}). 
The curve is in good agreement with the experimental result. 
In our model, 
the main reason for the degradation 
of the visibility is estimated to be the effect of the multi-photons 
in the coherent light pulse, and thus 
the demonstrated frequency down-converter 
is expected to have small adverse effects in the interference. 
If we replace the coherent light pulse 
by a heralded single photon prepared similarly to the mode B 
in Fig.~\ref{fig:setup}, 
the visibility of the two-photon interference 
will become $0.91$ as a result of the above analysis. 
Such a situation is more compatible 
with the conceptual setup in Fig.~\ref{fig:concept}. 

In conclusion, we have demonstrated 
the HOM interference of the two light pulses 
which are frequency down-converted to the telecommunication wavelength 
of 1522 nm from the visible wavelength of 780 nm. 
We observed 
a visibility of $0.76\pm 0.12$, 
which clearly shows 
the non-classical interference of the two light pulses. 
The theoretical analysis indicates that 
the degradation of the visibility is mainly caused 
by the multi-photons in the coherent light pulse, 
and the visibility expected when we use heralded single photons 
for the photon sources is $0.91$. 
We believe that the high-visibility wavelength conversion 
will be one of the key devices 
for many applications of quantum communication 
over long distance such as quantum repeaters. 

This work was supported by the Funding Program 
for World-Leading Innovative R \& D 
on Science and Technology~(FIRST), 
MEXT Grant-in-Aid for Scientific Research 
on Innovative Areas 20104003 and 21102008, 
the MEXT Global COE Program, 
and MEXT Grant-in-Aid for Young scientists(A) 23684035.


\begin{thebibliography}{999}

\section{References}

\bibitem{conv} 
P. Kumar, 
Opt. Lett. {\bf 15}, 1476 (1990). 

\bibitem{tanzilli} 
S. Tanzilli {\it et al.}
Nature~(London) {\bf 437}, 116 (2005). 

\bibitem{takesue}
H. Takesue, 
\prl {\bf 101}, 173901 (2008). 

\bibitem{matthew} 
M. T. Rakher, L. Ma, O. T. Slattery, X. Tang, 
and K. A. Srinivasan, 
Nature Photonics {\bf 4}, 786 (2010).

\bibitem{mcguinness}
H. J. McGuinness, M. G. Raymer, C. J. McKinstrie, and S. Radic, 
\prl {\bf 105}, 093604 (2010). 

\bibitem{dudin} 
Y. O. Dudin {\it et al.} 
\prl {\bf 105}, 260502 (2010). 

\bibitem{ikuta}
R. Ikuta {\it et al.}, 
Nature Commun. {\bf 2}, 537 (2011). 

\bibitem{ramelow}
S. Ramelow, A. Fedrizzi, A. Poppe, N. K. Langford, and A. Zeilinger, 
\pra {\bf 85}, 013845 (2012). 

\bibitem{ikuta2}
R. Ikuta {\it et al.}, 
\pra {\bf 87}, 010301(R) (2013). 

\bibitem{zaske}
S. Zaske {\it et al.}, 
\prl {\bf 109}, 147404 (2012). 

\bibitem{ates}
S. Ates {\it et al.}, 
\prl {\bf 109}, 147405 (2012). 

\bibitem{repeaters}
N. Sangouard, C. Simon, H. de Riedmatten, and N. Gisin, 
Rev. Mod. Phys {\bf 83}, 33 (2011). 

\bibitem{matsukevich}
D. N. Matsukevich, and A. Kuzmich, 
Science {\bf 306}, 663\,-\,666 (2004). 

\bibitem{rosenfeld}
W. Rosenfeld {\it et al.}, 
\prl {\bf 101}, 260403 (2008). 

\bibitem{chou}
C. W. Chou {\it et al.}, 
Nature~(London) {\bf 438}, 828 (2005). 

\bibitem{olmschenk}
S. Olmschenk {\it et al.}, 
Science {\bf 323}, 486 (2009). 

\bibitem{togan}
E. Togan {\it et al.}, 
Nature~(London) {\bf 466}, 730 (2010). 

\bibitem{ritter}
S. Ritter {\it et al.}, 
Nature~(London) {\bf 484}, 195 (2012). 

\bibitem{hofmann}
J. Hofmann {\it et al.}, 
Science {\bf 337}, 72 (2012). 

\bibitem{beugnon}
J. Beugnon {\it et al.}, 
Nature~(London) {\bf 440}, 779 (2006). 

\bibitem{maunz}
P. Maunz {\it et al.}, 
Nature Physics {\bf 3}, 538 (2007).

\bibitem{moehring}
D. L. Moehring {\it et al.}, 
Nature~(London) {\bf 449}, 68 (2007). 

\bibitem{bernien}
H. Bernien {\it et al.}, 
\prl {\bf 108}, 043604 (2012). 

\bibitem{sipahigil}
A. Sipahigil {\it et al.}, 
\prl {\bf 108}, 143601 (2012). 

\bibitem{HOM}
C. K. Hong, Z. Y. Ou, and L. Mandel, 
\prl {\bf 59}, 2044 (1987). 

\bibitem{Ou}
Z. -Y. J. Ou, {\it Multi-photon Quantum Interference} 
(Springer, New York, 2007). 

\bibitem{Rarity1}
J. G. Rarity, P. R. Tapster, and R. Loudon, 
arXiv: quant-ph/9702032 (1997). 

\bibitem{Rarity2}
J. G. Rarity, P. R. Tapster, and R. Loudon, 
J. Opt. B {\bf 7}, S171 (2005). 

\bibitem{Ruibo}
R. -B. Jin {\it et al.}, 
\pra {\bf 83}, 031805(R) (2011). 

\bibitem{nishikawa} 
T. Nishikawa {\it et al.}, 
Opt. Express {\bf 17}, 17792 (2009). 

\bibitem{NICT1}
S. Miki, M. Takeda, M. Fujiwara, M. Sasaki, and Z.Wang, 
Opt. Express {\bf 17}, 23558 (2009).

\bibitem{NICT2} 
S. Miki, T. Yamashita, M. Fujiwara, M. Sasaki, and Z. Wang, 
Opt. Lett. {\bf 35}, 2133 (2010). 

\bibitem{koashi}
M. Koashi, \prl {\bf 93}, 120501 (2004). 

\end{thebibliography}
\end{document}